\documentclass[twocolumn,showpacs,preprintnumbers,amsmath,amssymb,nofootinbib]{revtex4}
\input psfig.sty 
\usepackage[dvips]{color}
\usepackage{graphicx}
\usepackage{epsfig}

\begin{document}

\title{Observational tests for $\Lambda$(t)CDM cosmology}

\author{C. Pigozzo$^{1}$\footnote{cpigozzo@ufba.br}, M. A. Dantas$^2$\footnote{aldinez@on.br}, S. Carneiro$^{1,3}$\footnote{saulo.carneiro@pq.cnpq.br} and J. S. Alcaniz$^2$\footnote{alcaniz@on.br}}

\affiliation{ $^1$Instituto de F\'{\i}sica, Universidade Federal da Bahia, Salvador, BA, Brazil\\ $^2$Observat\'orio Nacional, Rio de Janeiro, RJ, Brazil\\ $^3$International Centre for Theoretical Physics, Trieste, Italy\footnote{Associate Member}}

\date{\today}

\begin{abstract}

We investigate the observational viability of a class of cosmological models in which the vacuum energy density decays linearly with the Hubble parameter, resulting in a production of cold dark matter particles at late times. Similarly to the flat $\Lambda$CDM case, there is only one free parameter to be adjusted by the data in this class of $\Lambda$(t)CDM scenarios, namely, the matter density parameter. To perform our analysis we use three of the most recent SNe Ia compilation sets (Union2, SDSS and Constitution) along with the current measurements of distance to the BAO peaks at $z = 0.2$ and $z = 0.35$ and the position of the first acoustic peak of the CMB power spectrum. We show that in terms of $\chi^2$ statistics both models provide good fits to the data and similar results. A quantitative analysis discussing the differences in parameter estimation due to SNe light-curve fitting methods (SALT2 and MLCS2k2) is studied  using the current SDSS and Constitution SNe Ia compilations. A matter power spectrum analysis using the 2dFGRS is also performed, providing a very good concordance with the constraints from the SDSS and Constitution MLCS2k2 data.

\end{abstract}


\maketitle

\section{Introduction}

The discovery of the late-time cosmic acceleration poses one of the greatest challenges theoretical physics has ever faced. In principle, this phenomenon may be the result of unknown physical processes involving either modifications of gravitation theory or the existence of new fields in high energy physics.  In the context of Einstein's general theory of relativity this means that some sort of dark energy, constant or that varies slowly with time and space, dominates the current composition of the cosmos (see, e.g., ~\cite{rev} for recent reviews).

Among the many possibilities to describe this exotic component, the simplest way is possibly by means of a cosmological constant $\Lambda$, which acts on the Einstein field equations as an isotropic and homogeneous source with $p_{\Lambda}= -\rho_{\Lambda}$. Historically, $\Lambda$ has been identified with the energy density of the quantum vacuum, but any attempt to obtain such contribution in quantum field theories in curved space-time leads to results many orders of magnitude above the observed value ($|\rho_{\Lambda}^{obs}| \simeq 10^{-10}$ $\rm{erg/cm^3}$), and even when an appropriate renormalization procedure is used, the resulting value usually disagrees with the actual one~\cite{weinberg}. This is the case, for example, of the vacuum density of free conformal fields in de Sitter space-time, which is shown to be of the order of $H^4$, where $H$ is the Hubble parameter \cite{H4}. Although such a result can be used to construct non-singular scenarios in the high-energy limit \cite{award}, at late-times it leads to a very tiny $\Lambda$.

A possible way out of this problem is to consider the contribution of interacting fields in the low-energy limit, which is a more realistic situation. In this sense, some authors have argued that the QCD chiral phase transition originates a vacuum energy density scaling as $\Lambda \approx m^3 H$, where $m \approx 150$ MeV is the energy scale of the QCD vacuum transition \cite{QCD}. By taking $\Lambda \sim H^2$ in the present Universe, we obtain $\Lambda \sim m^6$, which coincides with the current order of magnitude of the cosmological term. Clearly, no theoretical result in this field is free of some assumptions, and a number of other proposals for a possible time dependence of $\Lambda$ can be found in the literature (see, e.g.,~\cite{Lambda(t)}). In particular, in the case of a vacuum density decaying linearly with $H$, one can show that the corresponding cosmological model is indistinguishable from the $\Lambda$CDM model during the phase dominated by radiation, but it departs from the standard scenario during the matter epoch, in part owing to the production of dark matter associated to the vacuum decay \cite{Borges}.

In a previous communication~\cite{anterior}, we have tested the observational viability of this class of $\Lambda(t)$CDM scenarios discussed above from a joint analysis involving type Ia supernovae data  (Supernova Legacy Survey Collaboration~\cite{snls}) along with the position of the first acoustic peak in the CMB spectrum of anisotropies ($l_1$) from WMAP~\cite{WMAP}, and the SDSS observed distance to the baryonic acoustic oscillation (BAO) peak at $z = 0.35$~\cite{Eisenstein}. The goal of the present contribution is to update the results of Ref.~\cite{anterior} by using three of the most recent SNe Ia data sets, namely, the updated compilation of Union sample (Union2)~\cite{union}, the  nearby + SDSS + ESSENCE + SNLS + Hubble Space Telescope (HST) set of 288 SNe Ia discussed in Ref.~\cite{sdss} (throughout this paper we refer to this set as SDSS compilation) and the Constitution set of 397 SNe
Ia \cite{cs}. We consider two sub-samples of these two latter compilations that use SALT2~\cite{salt2} and MLCS2k2~\cite{mlcs2k2} SNe Ia light-curve fitting methods. Along with the SNe Ia data, and to improve the bounds on the free parameter of the model we use, besides the current value of $l_1$ and the BAO peak at $z = 0.35$, the distance to the BAO peak at $z = 0.2$, as recently discussed in Ref.~\cite{percival}. For the sake of comparison we also perform the same analyses for the standard $\Lambda$CDM model.

In order to verify the consistence between background and perturbative tests, we also perform a joint analysis including the observed matter power spectrum from the 2dFGRS.  In the case of SDSS (MLCS2k2) and Constitution (MLCS2k2 17) data, we obtain a very good concordance with $\Lambda(t)$ model. Since MLCS2k2 depends weakly on standard cosmology as compared to SALT2, it is actually the most appropriate to test non-standard models, which suggests that the obtained concordance is robust.

This paper is organized as follows. In Sec. II we discuss the basic equations and some features of the vacuum decay model with $\Lambda \propto H$. Observational tests involving SNe Ia, BAO, CMB and LSS data are performed in Sec. III. In Sec. IV we present our main results. We also discuss quantitatively the influence of two SNe Ia light-curve fitting methods on parameter estimates using the current SDSS and Constitution compilation sets. We end by summarizing our main conclusion in Sec. V.

\section{Basics of the $\Lambda(t)$ model}

In units where $(8\pi G)^{1/2} = c = 1$, the Friedmann equations for a spatially flat universe can be written as 
\begin{equation} \label{friedmann}
\rho_T = 3H^2\;,
\end{equation}
\begin{equation}
\dot{\rho}_T + 3H(\rho_T + p_T) = 0\;,
\end{equation}
where $\rho_T$ and $p_T$ are the total energy density and pressure, respectively, and the second equation expresses the conservation of the total energy. By taking the cosmic fluid as composed by dust matter of density $\rho_m$ and a time-dependent cosmological term with $p_{\Lambda} = - \Lambda$, we have
\begin{equation} \label{continuidade}
\dot{\rho}_m + 3H\rho_m = -\dot{\Lambda}\;,
\end{equation}
which means that the decay of the vacuum energy density is concomitant with matter production. In order to avoid problems with primordial nucleosynthesis and the spectrum of CMB, we must postulate that no baryonic matter or radiation is produced in this process. Therefore, the present model may be considered as a particular case of models with interaction only in the dark sector (see, e.g., \cite{ernandes} and Refs. therein for other examples of scenarios with interaction in the dark sector).

With the ansatz $\Lambda = \sigma H$, where $\sigma$ is a constant of the order of $m^3$, the solution of the above equations is given by \cite{Borges}
\begin{equation} \label{a}
a(t) = C \left[\exp\left(\sigma t/2\right) - 1\right]^{\frac{2}{3}}\;,
\end{equation}
which coincides with the Einstein-de Sitter solution in the limit of early times and at late times tends asymptotically to the de Sitter space-time.

The corresponding Hubble parameter as a function of the redshift is given by \cite{primeiro,anterior}
\begin{equation} \label{Hz}
H(z) = H_0 \left[1 - \Omega_{m} +\Omega_{m} (1 + z)^{3/2}\right],
\end{equation}
where $\Omega_m$ and $H_0 = 100h$ $\rm{km/s/Mpc}$ are, respectively, the current values of the matter density and Hubble parameters. Note that, similarly to the flat $\Lambda$CDM scenario, these are the only two parameters of the model to be adjusted by the data. Note also that this class of $\Lambda$(t)CDM cosmologies does not contain the $\Lambda$CDM scenario as a particular case, which means that in principle they may be observationally distinguishable.

\section{Observational tests}

In this Section we investigate some observational constraints on both models from five different SNe Ia data sets along with the current value of $l_1$ from WMAP, the distance to the BAO peak at $z = 0.2$ and $z = 0.35$ and data of Large Scale Structure (LSS) distribution from 2dFGRS.

\subsection{SNe Ia data}

We use in our analyses five of the most recent SNe Ia compilations available, namely, the Union2 sample of Ref.~\cite{union}, the two compilations of the SDSS collaborations discussed in Ref.~\cite{sdss} and two of the Constitution compilations considered in \cite{cs}\footnote{In \cite{cs} the name Constitution refers to data calibrated with SALT. Here we will use it to refer to any compilation presented there.}.

The Union2 sample\footnote{http://www.supernova.lbl.gov/} is an update of the original Union compilation. It comprises 557 data points including recent large samples from other surveys and uses SALT2 for SN Ia light-curve fitting. The SDSS compilation\footnote{http://www.sdss.org/} of 288 SNe Ia uses both SALT2 and MLCS2k2 light-curve fitters\footnote{It is worth emphasizing that SALT2 fitter does not provide a cosmology-independent distance estimate for each SNe, since some parameters in the calibration process are determined in a simultaneous fit with cosmological parameters to the Hubble diagram (see~\cite{friemanON} for a discussion). See also \cite{36} for a discussion on possible systematic biases from the multicolor light curve shape method.} and is distributed in redshift interval $0.02 \leq z \leq 1.55$. This sample, along with the so-called CMB/BAO ratio, has been used in Ref.~\cite{sollerman} to perform a comparative analysis involving a number of nonstandard cosmological scenarios (we refer the reader to this latter reference for more on the SDSS compilation). The Constitution compilations comprises 397 SNe Ia (CfA3 + Union) calibrated with SALT2 and two versions of the MLCS2k2 fitter: MLCS31 and MLCS17. As discussed in \cite{cs}, the former overestimates host-galaxy extinction, and here we will consider only the later version.

\subsection{BAO}

The second observable we consider is the distance to a given observed baryonic acoustic oscillation, which is defined as
\begin{equation} \label{DV}
D_V(z) = \left[D_M(z)^2 \frac{z}{H(z)}\right]^{1/3}\;,
\end{equation}
where $D_M$ is the comoving angular-diameter distance. We will consider two different measurements of BAO, at redshifts $z_{\rm{BAO}} = 0.20$ and $z_{\rm{BAO}} = 0.35$, i.e.,
$$
D_V(z = 0.2) = 748 \pm 57\; {\rm{Mpc}},
$$
$$
D_V(z = {0.35}) = 1300 \pm 88\; {\rm{Mpc}},
$$
where the value of $D_V$ at $z = 0.2$ is obtained by combining the results of Ref.~\cite{percival} with the value of $D_V(z = 0.35)$ obtained in Ref.~\cite{tegmark}.

An important aspect worth emphasizing at this point is that we cannot use in our analysis the parameter ${\cal{A}}$, defined as ${\cal{A}} = D_V(z)\sqrt{\Omega_mH_0^2}/z_{\rm{BAO}}$ \cite{Eisenstein}, since in the present model the production of matter leads to a different dependence of the horizon at the time of matter-radiation equality on the present matter density \cite{anterior}. 
In all the cases, the observable quantity is the ratio $r_s(z_d)/D_V$, where $r_s(z_d)$ is the sound horizon at the time of drag epoch. Since $r_s(z_d)$ is very weakly model-dependent (provided the radiation epoch is the same as in the $\Lambda$CDM model), the distance $D_V$ can in fact be used in our analysis. As well known, it is also obtained from direct observations by using the $\Lambda$CDM as a fiducial model~\cite{Eisenstein}, but such a dependence is weak, and actually this particular observable does not impose strong limits on the model parameter~\cite{anterior}.

\subsection{CMB}

The most restrictive of the background tests we will perform in our analysis is the position of the first peak of the CMB spectrum of anisotropies. The simplest way to use it is certainly by considering the shift parameter, defined as ${\cal{R}} = \sqrt{\Omega_m}D_M(z*)$, where $z_*$ is the redshift of the last-scattering surface. But again, this is not allowed in our analysis~\cite{anterior}, since the angular scale $l_A$ is not the same as in the $\Lambda$CDM case\footnote{For the same reason, we cannot use in the joint analysis the ratio $l_A/D_V$, used in other analyses.}. We, therefore, use directly the observed position of the peak, $l_1 = 220.8 \pm 0.7$~\cite{WMAP}, which is related to the angular scale by~\cite{tegmark1}
\begin{equation}\label{l1}
l_1 = l_A (1 - \delta_1), \quad  \mbox{where} \quad \delta_1 = 0.267 \left( \frac{r}{0.3} \right)^{0.1}
\end{equation}
with $r \equiv \rho_{r}(z_{ls})/\rho_m(z_{ls})$. Here, $\rho_{r}(z_{ls})$ is the radiation density at the time of last scattering\footnote{As shown in \cite{anterior}, in the present model the redshift of last scattering is the same as in the $\Lambda$CDM case.}.
Properly speaking, these relations were obtained for the $\Lambda$CDM case, but they depend only on pre-recombination physics and can be applied to the present model as a good approximation. Nevertheless, owing to the process of matter production, the matter density at the time of last scattering is $\rho_m = 3H_0^2\Omega_m^2 z_{ls}^3$, i.e., smaller than the $\Lambda$CDM value by a factor $\Omega_m$, what can be seen directly from the high-$z$ limit of the Hubble function (\ref{Hz}). Therefore, in the present case we have~\cite{anterior}
\begin{equation}
r = \frac{\Omega_{r}}{\Omega_m^2} z_{ls},
\end{equation}
leading to a relation between $l_A$ and $l_1$ different from the $\Lambda$CDM case.

The theoretical expression for the acoustic scale is defined as the ratio
\begin{equation}\label{lA}
l_A = {\pi \int_0^{z_{ls}}
\frac{dz}{H(z)}}/{\int_{z_{ls}}^{\infty}\frac{c_s}{c}\frac{dz}{H(z)}},
\end{equation}
with the sound velocity given by
$c_s = c \left( 3 + \frac{9}{4}\frac{\Omega_b}{\Omega_{\gamma}z}
\right)^{-1/2}$,
where $\Omega_b$ and $\Omega_{\gamma}$ stand for the present density parameters of baryons and photons, respectively. The Hubble function used in (\ref{lA}) must take into account the radiation contribution $\Omega_r$ and is rewritten as~\cite{anterior}
\begin{equation}\label{Hgeral}
\frac{H(z)}{H_0} \approx \left\{ \left[1 - \Omega_{m} +\Omega_{m} (1
+ z)^{3/2}\right]^2 + \Omega_r (1+z)^4 \right\}^{1/2}\;.
\end{equation}

\subsection{The matter power spectrum}

Except for scales near the horizon,  vacuum perturbations are negligible (see \cite{Julio,Zimdahl}). Therefore, we simplify our analysis by considering an unperturbed time-dependent cosmological term. The model prediction (blue line) together with the 2dFGRS data points~\cite{2dF} are shown in  the left panel of Figure 6. For comparison, the concordance $\Lambda$CDM fitting as provided by the BBKS transfer function~\cite{BBKS} is also shown (red line), as well as the result for the $\Lambda$CDM case of the same simplified analysis we have performed with $\Lambda$(t)CDM (black line). In such analyses baryons are not included, which leads to an error of about $10\%$ as compared to a more exact study \cite{Julio}. In the case of $\Lambda$CDM, the best fit is given by $\Omega_m \approx 0.2$, in agreement with \cite{2dF}.

\begin{figure*}
\vspace{.2in}
\centerline{\psfig{figure=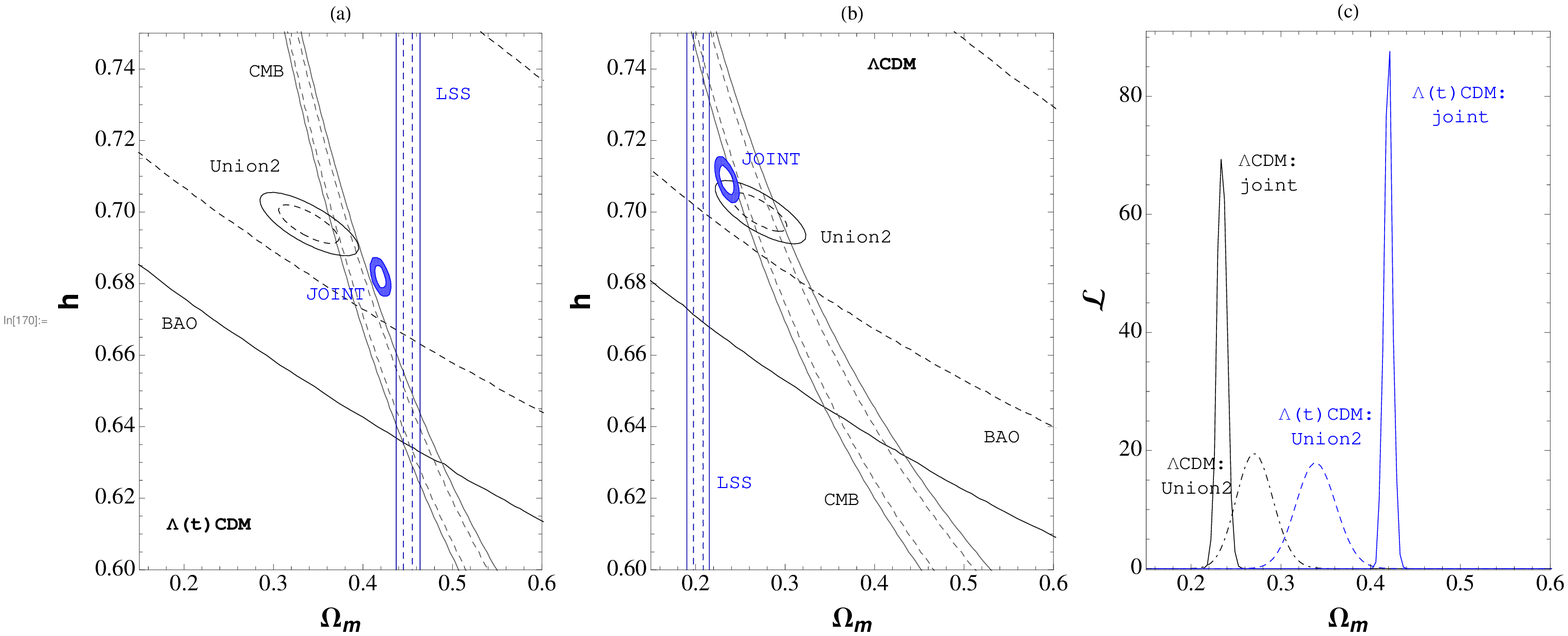,width=7.1truein,height=3.1truein}
\hskip 0.1in}
\caption{{\bf{a)}} Superposition of the confidence regions in the $h \times \Omega_m$ plane from Union2 SNe Ia, CMB, BAO and LSS data sets discussed in the text, for the $\Lambda$(t)CDM model. The blue elipse shows the joint $2\sigma$ confidence region.  {\bf{b)}} The same as in Panel {\bf{(a)}} for the $\Lambda$CDM model. {\bf{c)}} The likelihood  ${\cal{L}}$ for $\Omega_m$ arising from Union2 SNe Ia sample for $\Lambda$(t)CDM (dot-dashed blue curve) and $\Lambda$CDM (dot-dashed  black curve) scenarios. The solid lines correspond to the joint analysis including CMB, BAO and LSS data sets.}
\end{figure*}

\begin{figure*}
\vspace{.2in}
\centerline{\psfig{figure=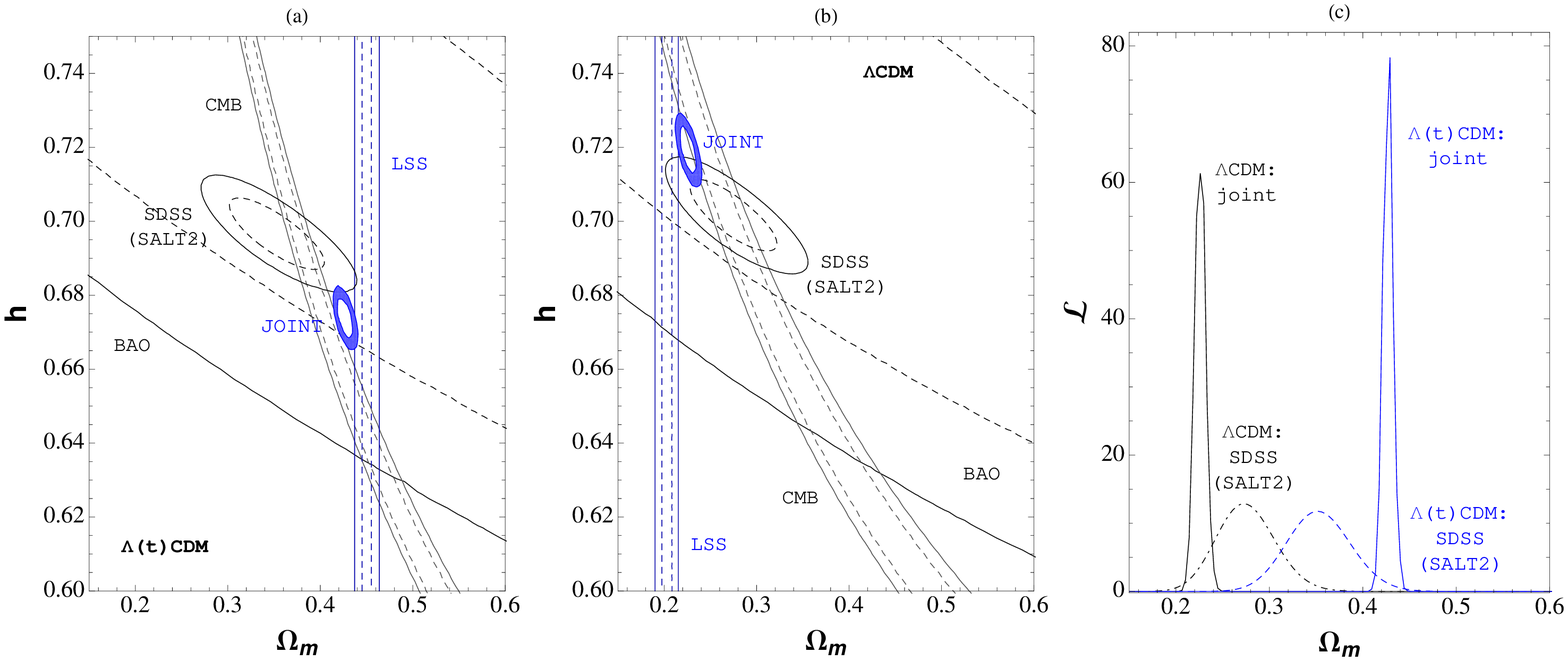,width=7.1truein,height=3.1truein}
\hskip 0.1in}
\caption{The same as in Figure 1 for SDSS SNe Ia data with SALT2 light-curve fitter.}
\end{figure*}

\begin{figure*}
\vspace{.2in}
\centerline{\psfig{figure=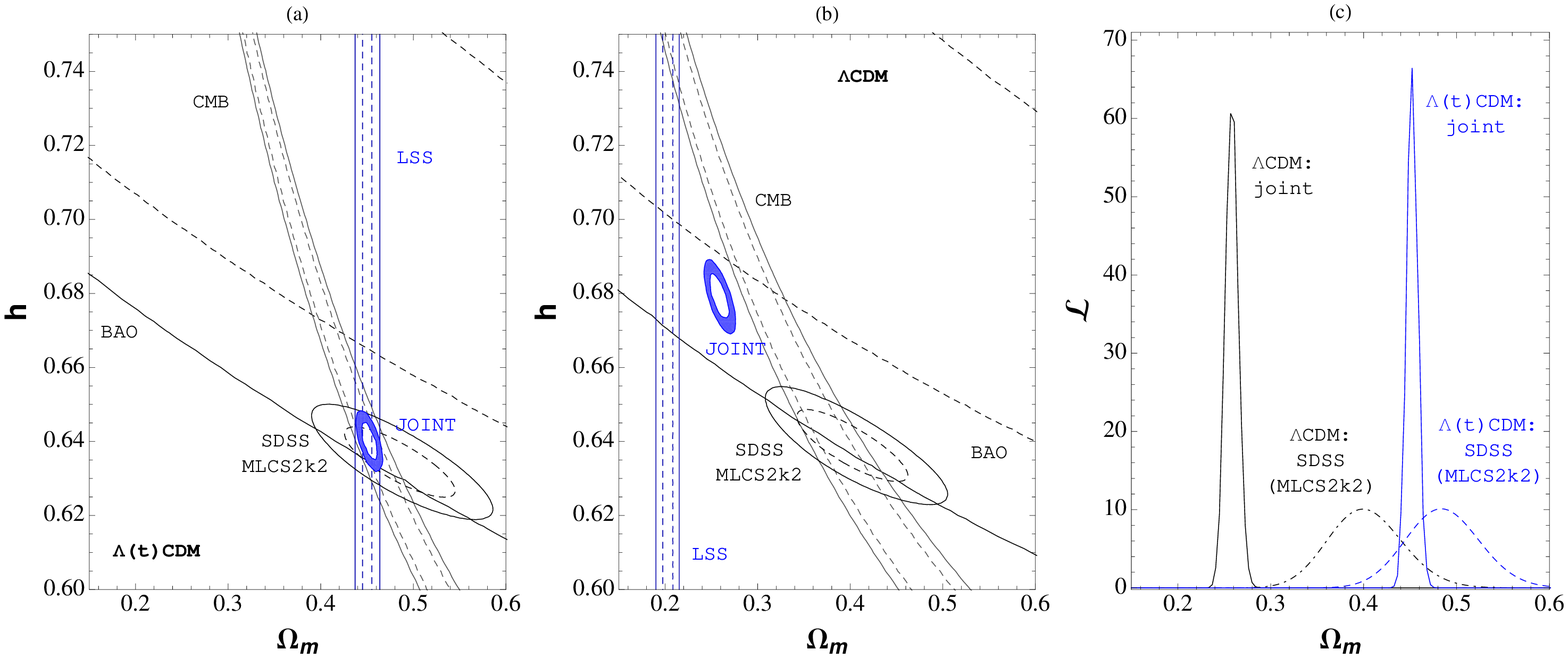,width=7.1truein,height=3.1truein}
\hskip 0.1in}
\caption{The same as in Figure 1 for SDSS SNe Ia data with MLCS2k2 light-curve fitter. By comparing Figs. 2 and 3 we clearly see the influence of SNe Ia light-curve fitting on parameter estimation.}
\end{figure*}

\begin{figure*}
\vspace{.2in}
\centerline{\psfig{figure=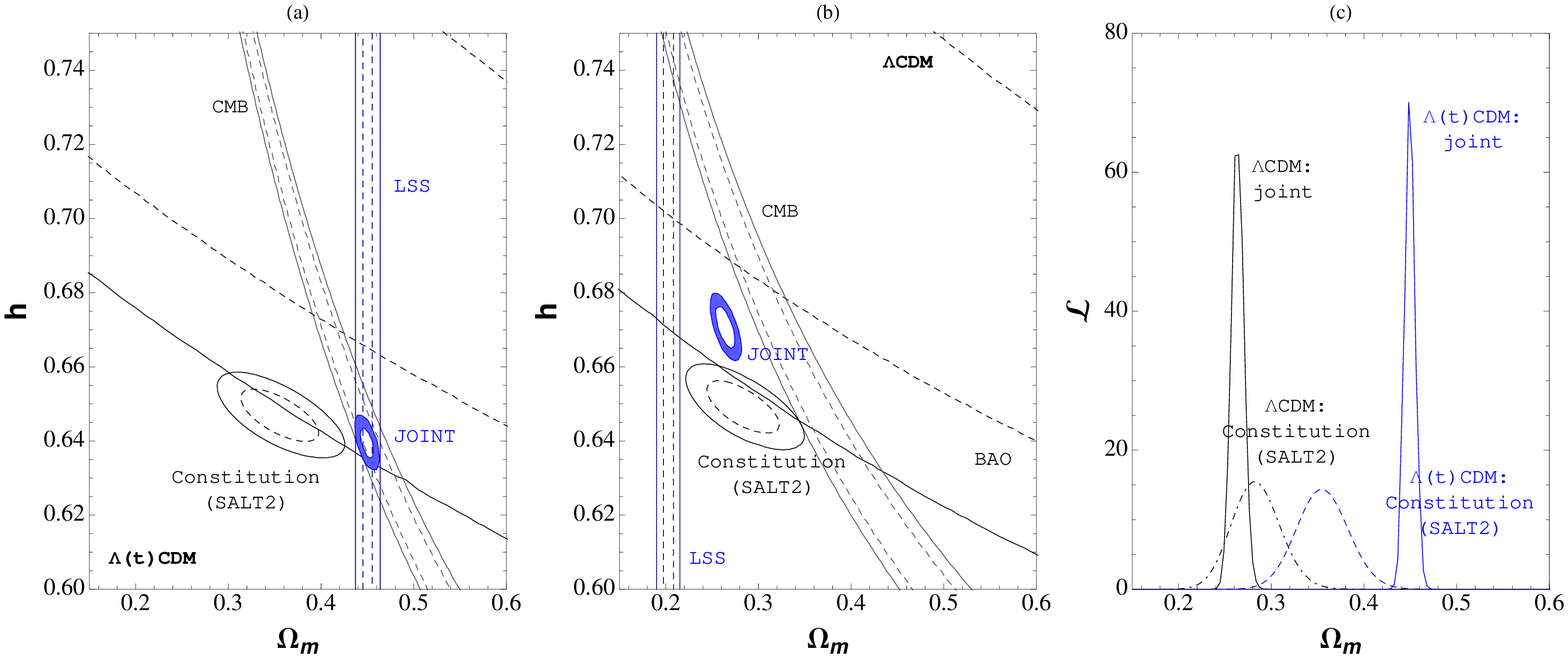,width=7.1truein,height=3.1truein}
\hskip 0.1in}
\caption{The same as in Figure 1 for Constitution SNe Ia data with SALT2 light-curve fitter.}
\end{figure*}

\begin{figure*}
\vspace{.2in}
\centerline{\psfig{figure=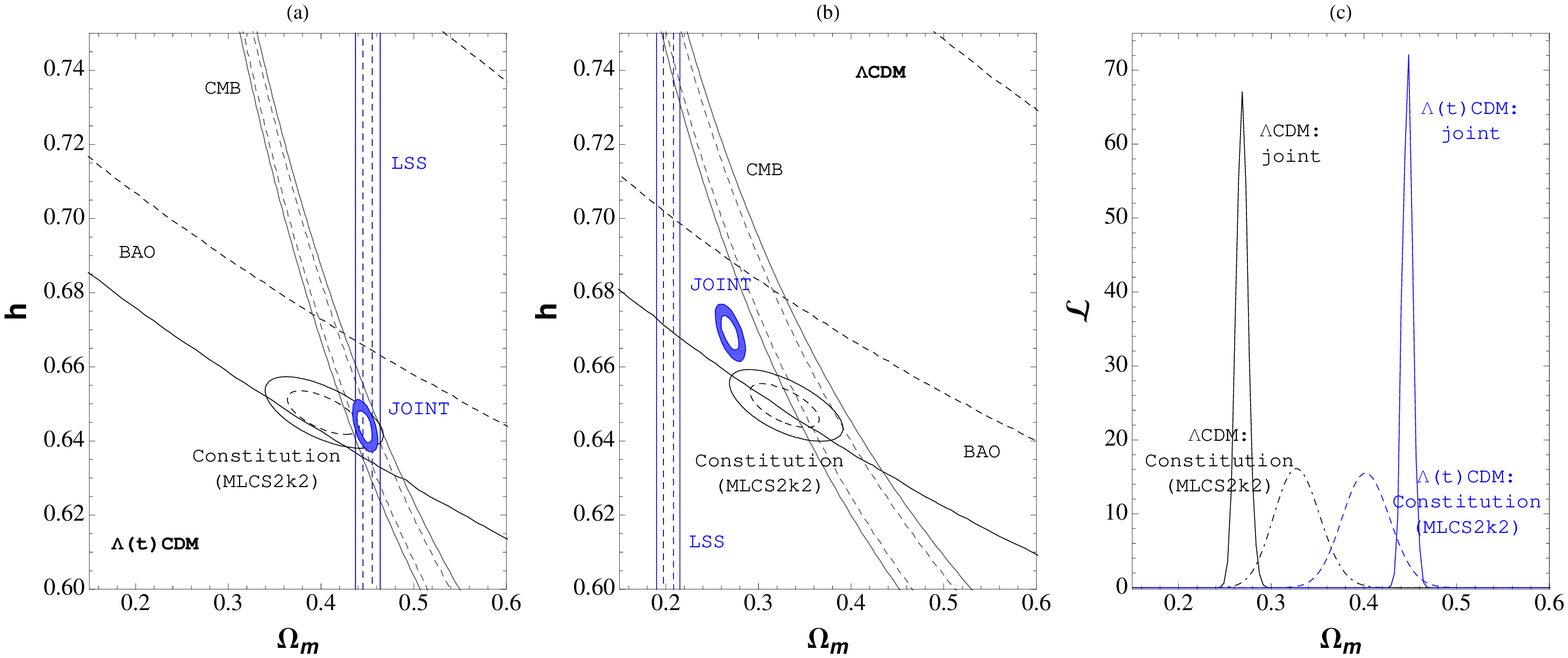,width=7.1truein,height=3.1truein}
\hskip 0.1in}
\caption{The same as in Figure 1 for Constitution SNe Ia data with MLCS17 light-curve fitter.}
\end{figure*}

\begin{table}[t]
\begin{center}
\caption{Limits to $\Omega_m$ ($\Lambda$(t)CDM).}
\begin{tabular}{rrll}
\hline  \hline \\
\multicolumn{1}{c}{Test}&
\multicolumn{1}{c}{$\Omega_m$\footnote{Error bars stand for $2\sigma$}}&
\multicolumn{1}{c}{$\chi^2_{min}$} &
\multicolumn{1}{c}{$\chi^2_{min}/\nu$}\\   \hline \\
Union2(SALT2)........& $0.339^{+0.045}_{-0.043}$ & 554.9 & 0.998\\
Union2(SALT2)\footnote{+ CMB + BAO+LSS}......&$0.420^{+0.009}_{-0.010}$ &634.5& 1.063\\
SDSS (SALT2).......&$0.352^{+0.071}_{-0.065}$& 256.9 &0.895\\
SDSS (SALT2)$^b$.......&$0.430^{+0.008}_{-0.014}$& 320.8 &0.978\\
SDSS (MLCS2k2).......& $0.484^{+0.082}_{-0.075}$ & 249.7 &0.870\\
SDSS (MLCS2k2)$^b$........&$0.450^{+0.014}_{-0.010}$& 276.3 &0.842\\
Constitution (SALT2).......& $0.355^{+0.056}_{-0.054} $& 343.3 &0.984\\
Constitution (SALT2)$^b$.......& $0.450^{+0.011}_{-0.011}$& 390.2 &0.998\\
Constitution (MLCS2k2).......& $0.402^{+0.052}_{-0.049} $& 397.2 &1.074\\
Constitution (MLCS2k2)$^b$.......& $0.450^{+0.008}_{-0.014}$& 435.3 &1.057\\
\hline  \hline
\end{tabular}
\end{center}
\end{table}

\begin{table}[t]
\begin{center}
\caption{Limits to $\Omega_m$ ($\Lambda$CDM)}
\begin{tabular}{rrll}
\hline  \hline \\
\multicolumn{1}{c}{Test}&
\multicolumn{1}{c}{$\Omega_m$\footnote{Error bars stand for $2\sigma$}}&
\multicolumn{1}{c}{$\chi^2_{min}$} &
\multicolumn{1}{c}{$\chi^2_{min}/\nu$}\\   \hline \\
Union2(SALT2)........& $0.270^{+0.043}_{-0.038}$& 552.7 &0.994\\
Union2(SALT2)\footnote{+ CMB + BAO+LSS}......&$0.235\pm0.011$& 613.1 &1.027\\
SDSS (SALT2).......&$0.273^{+0.067}_{-0.057}$ & 256.0 &0.892\\
SDSS (SALT2)$^b$.......&$0.226\pm0.012$ & 305.8 &0.932\\
SDSS (MLCS2k2).......&$0.399^{+0.087}_{-0.072}$& 249.1 &0.868\\
SDSS (MLCS2k2)$^b$........&$0.260^{+0.013}_{-0.016}$ & 403.8 &1.231\\
Constitution (SALT2).......& $0.282^{+0.054}_{-0.049} $ & 341.7 &0.979\\
Constitution (SALT2)$^b$.......& $0.265^{+0.013}_{-0.014}$ & 515.8 &1.319\\
Constitution (MLCS2k2).......& $0.327^{+0.052}_{-0.047} $ & 397.4 &1.074\\
Constitution (MLCS2k2)$^b$.......& $0.270\pm0.013$ & 570.2 &1.384\\
\hline  \hline
\end{tabular}
\end{center}
\end{table}

\section{Analysis and results}

In order to derive the constraints on our model we maximaze the likelihood function ${\cal{L}} \propto \exp{(-\chi^2/2)}$, where $\chi^2 = \chi^2_{\rm{SNe}} + \chi^2_{\rm{BAO}} + \chi^2_{\rm{CMB}} + \chi^2_{\rm{LSS}}$ takes into account all the data sets discussed above. In our analyses, the Hubble parameter $H_0$ is considered as a nuisanse parameter so that we marginalize over it. The results are shown in Figs. 1-5, and Tables I and II summarize the main results of this paper for all SNe data sets and joint analyses. In order to compare the two classes of models discussed in this paper we also show the values of $\chi^2_{min}$ and $\chi^2_{min}/\nu$ ($\nu$ stands for degrees of freedom) for each analysis performed. 

In the figures, the left and central panels show the superposition of the $1\sigma$ and $2\sigma$ confidence regions in the $h - \Omega_m$ plane for the four tests we are considering, as well as the concordance ellipses of our joint analyses. In the case of SDSS (MLCS2k2), the interval for the matter density in $\Lambda$(t)CDM is about $10\%$ higher than the flat $\Lambda$CDM value. Both values, on the other hand, are higher than the usually accepted concordance value $\Omega_m \approx 0.27$ ~\cite{WMAP}. We also note that in both cases the value of $H_0$ is relatively small ($h \simeq 0.64$) compared to current estimates~\cite{h}. Nevertheless, when we marginalize over $h$ the matter density remains unchanged in both models. The right panels of the figures show the likelihood for $\Omega_m$ after marginalization.  In all the analyses the estimate of the matter density parameter for the $\Lambda$(t)CDM scenario is considerably larger than the corresponding value for the $\Lambda$CDM model. In particular, the estimates of $\Omega_m$ for the vacuum decay model is up to $\sim 2\sigma$ off from the central value inferred by applying the virial theorem to cluster dynamics~\cite{carlberg} (see also \cite{dekel}), but only $\sim 0.5\sigma$ off from  the central value obtained by using other independent methods, such as the mean relative peculiar velocity measurements for pairs of galaxies~\cite{feldman}.

In terms of parameter estimation we note that very similar results are found for both Union2 and SDSS compilation with SALT2.  An interesting aspect worth emphasizing is realized when a comparison between the analyses of Figs. 2 and 3 is made. Note that, although the only difference between these two analyses is the light-curve fitter used in the SNe calibration, the estimated values of the cosmological parameter $\Omega_m$ are considerably modified from analysis to analysis. For instance, by comparing only the SNe Ia bounds on $\Omega_m$ for the standard $\Lambda$CDM model, we observe that the estimated values of the matter density parameter increases $\sim 47\%$, from 0.27 to 0.40 (see Table II). The same can be said about the Constitution sample calibrated with SALT2 and MLCS17.  Similar conclusions are also drawn for the class of vacuum decay models studied here. In this case, estimates of the matter density parameter increase $\sim 38\%$ for the SDSS SNe Ia data with MLCS2k2 light-curve fitter, in better accordance with the independent estimate based on large scale structure distribution \cite{Julio}. It is worth observing that these estimated values of $\Omega_m$ obtained from SDSS (MLCS2k2) analyses for both models are relatively large, being $\sim 2\sigma$ off from the central value obtained using the cluster dynamics methods discussed earlier.

In all the cases above we have considered spatially flat scenarios, usually predicted by inflation.
For completness, we can relax this prior and include spatial curvature in both models. In the right panel of Figure 6 we show the results for the SDSS (MLCS2k2) data set \cite{Cassio}. We can see that the spatially flat universe is in accordance with data within $1\sigma$ confidence level in both models.

\begin{figure*}[]
\vspace{.2in}
\centerline{\psfig{figure=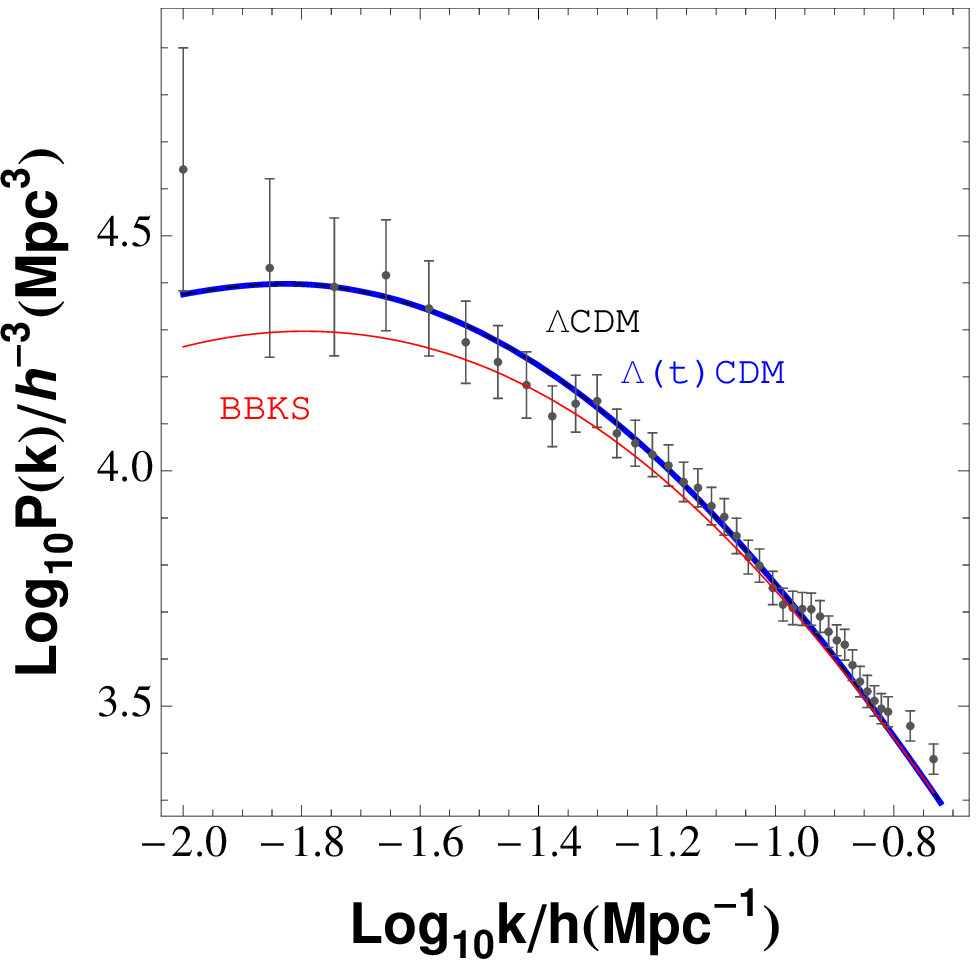,width=3.0truein,height=3.0truein}
\hspace{1.3cm}
\psfig{figure=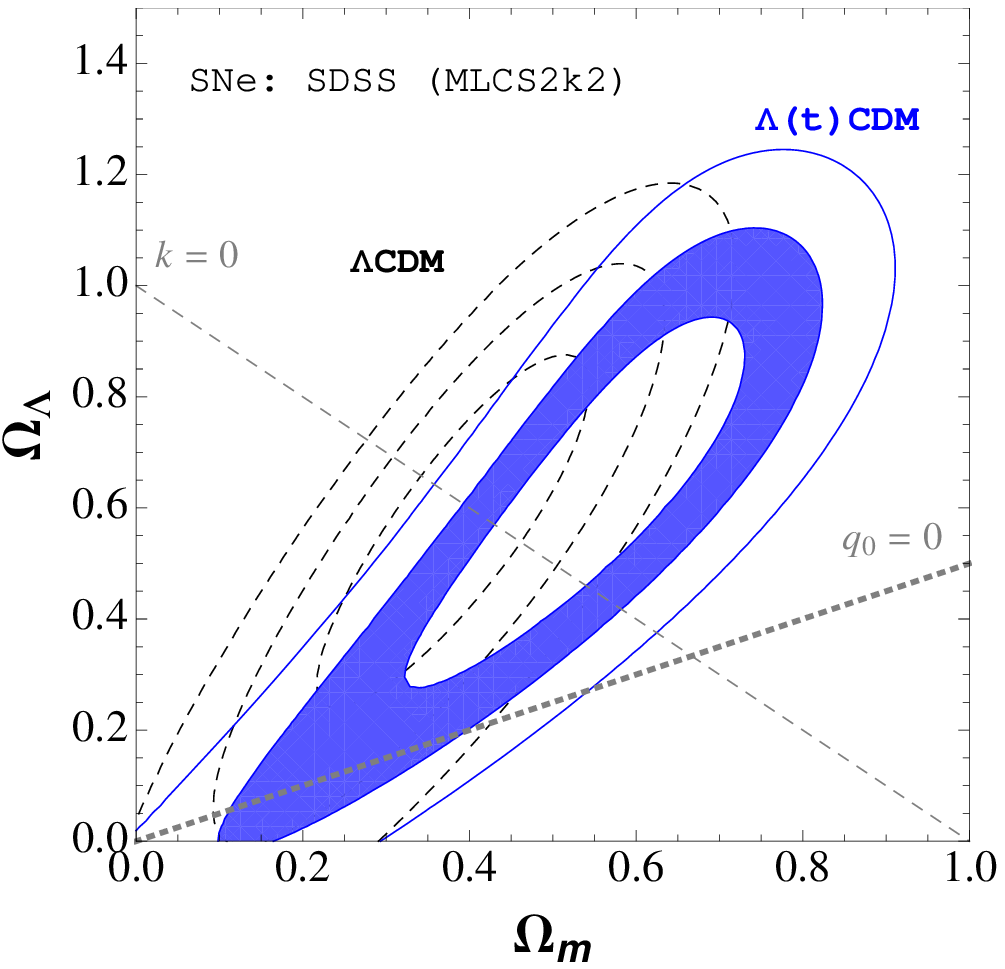,width=3.0truein,height=3.0truein}}
\caption{{\bf Left.} The theoretical power spectrum for the $\Lambda$(t)CDM model (blue line), for the simplified (without barions)  $\Lambda$CDM model (black line) and for the BBKS transfer function (red line). The data points are taken from \cite{2dF} (2dFGRS). {\bf Right.} The confidence regions in the $\Omega_{\Lambda} \times \Omega_m$ plane for the $\Lambda$CDM and $\Lambda$(t)CDM models with spatial curvature, when we use the SDSS (MLCS2k2) SNe Ia compilation. The lines labled $k=0$ and $q_0=0$ correspond, respectively, to spatially flat and uniformely expanding universes.}
\end{figure*}

\section{Final remarks}

Although in good agreement with current observations, a residual $\Lambda$ term exacerbates the well known cosmological constant problem, requiring a natural explanation for its small, but nonzero, value. In this paper, we have discussed the observational viability of a class of alternative $\Lambda(t)$CDM scenarios in which the vacuum energy density decays linearly with the Hubble parameter $H$. By considering the most recent SNe Ia compilations sets along with the current measurements of distance to the BAO peaks at $z = 0.2$ and $z = 0.35$ and the position of the first acoustic peak of the CMB power spectrum, we show that in terms of $\chi^2$ statistics both models provide good fits to the data with the very same number of free parameters (see Tables I and II). A recurrent aspect in all the analyses performed is the larger estimates of $\Omega_m$ for $\Lambda(t)$CDM relative to the values obtained in the $\Lambda$CDM context. When compared with some $\Omega_m$ estimates from independent observational methods~\cite{carlberg,dekel,feldman}, these values may be off from $\sim 0.5\sigma$ up to $\sim 2\sigma$. The same conclusion also arises in the $\Lambda$CDM context for the tests including SDSS (MLCS2k2).

We have also discussed quantitatively differences in parameter estimates between SALT2 and MLCS2k2 light-curve fitters using the current SDSS and Constitution SNe Ia compilations. We have shown that for both models these differences may reach up to $\sim 47\%$ for the matter density parameter. These results reinforce the need and importance for a better understanding of unresolved systematic effects on SNe Ia  calibration, as has been discussed by other authors (see, e.g., \cite{sollerman}).

Finally, from a joint analysis which includes the observed matter power spectrum, it is clear that a good concordance is achived for the $\Lambda$(t)CDM model when we use the SDSS (MLCS2k2) and Constitution (MLCS17) compilations of SNe Ia. Since the MLCS2k2 fitter depends weakly on standard cosmology, it is generally considered the most appropriate to test alternative models, which means that the concordance obtained here can be considered robust. Actually, an inspection of panels (c) of Figures 1-5 shows that, in the case of MLCS2k2 fitter, the $\Lambda(t)$ model presents better concordance than the standard model.

\section*{Acknowledgements}

The authors acknowledge CNPq - Brazil for the grants under which this work was carried out. JSA  is grateful for the support of CNPq grants 304569/2007-0 and  481784/2008-0. SC is thankful for the CNPq grant 305133/2008-0.

{}

\end{document}